\documentclass[12pt,a4paper]{article}
\usepackage{graphicx}
\usepackage{a4wide}
\usepackage{amsmath}

\begin{document}
\thispagestyle{empty}

\title{Spherical spaces for cosmic topology and multipole selection rules.}
\author{Peter Kramer,\\ Institut f\"ur Theoretische Physik der Universit\"at  T\"ubingen,\\ Germany.}
\maketitle

\section*{Abstract.}
Spherical manifolds yield cosmic spaces with positive curvature. 
They result by closing pieces from the sphere used by Einstein for his initial cosmology.
Harmonic analysis  on the manifolds  aims at  explaining 
the observed low amplitudes at small multipole orders of the cosmic microwave background. We analyze assumptions of point symmetry  and randomness  for spherical spaces. There emerge four spaces named orbifolds, with low volume fraction from the sphere  and sharp multipole selection rules in their eigenmodes.

\begin{figure}[tb]
\begin{center}
\includegraphics[width=0.8\textwidth]{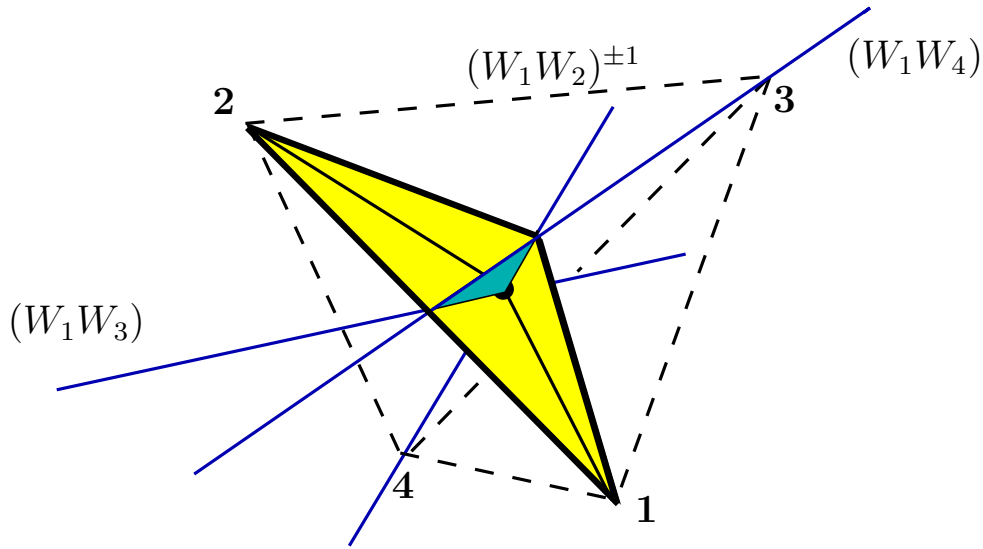} 
\end{center}
\caption{\label{fig:Fig1}
{\bf Tetrahedral orbifold.} Face gluing deck rotations of the tetrahedral  duplex orbifold (yellow) $N8$ in Euclidean version. 
The vertices of the Platonic tetrahedron are marked with numbers $(1, 2, 3, 4)$.
The four covering  rotations  $(W_1W_3), 
(W_1W_2)^{\pm 1},(W_1W_4)$ of the orbifold as products of Weyl reflections from $\Gamma =\circ - \circ -\circ -\circ$ are marked 
by rotations axes, intersection lines of Weyl planes. The three intersections of these lines with the orbifold
form the glue triangle (light blue area). Its inner edge points  have  the orders (2, 3, 2) 
of their covering rotations.}
\end{figure}

\begin{figure}[tb]
\begin{center}
\includegraphics[width=0.9\textwidth]{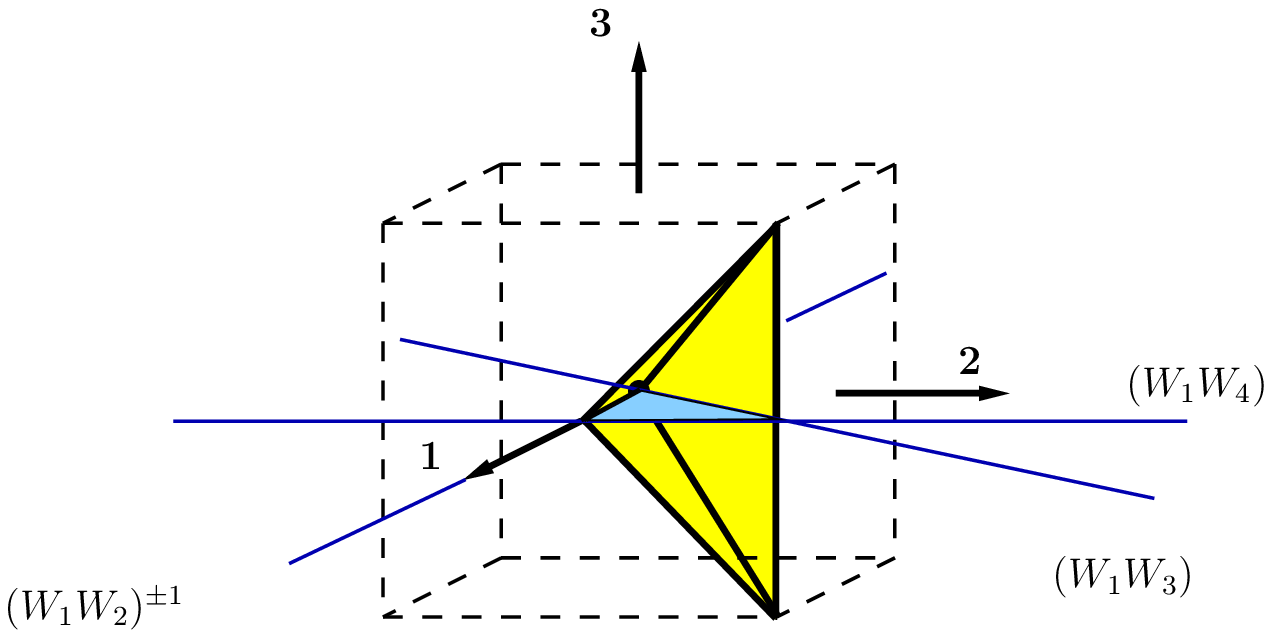} 
\end{center}
\caption{\label{fig:Fig2}
{\bf Cubic orbifold}. Face gluing deck rotations of the cubic duplex orbifold (yellow) $N9$ in Euclidean version. Its four covering rotations  $(W_1W_3),  
(W_1W_2)^{\pm 1}, (W_1W_4)$ are products of two  Weyl reflections from $\Gamma =\circ \stackrel{4}{-} \circ -\circ -\circ$ and marked 
by rotation axes, intersection lines of their Weyl planes. The three intersections of these lines with the orbifold
form the glue  triangle (light blue area). Its inner edge points have the orders (2, 4, 2) 
of their covering rotations.}
\end{figure}

\section{Introduction.}
\label{Int}

Einstein in his initial cosmology \cite{EI17}, see von Laue  \cite{LA56} pp. 160-4  and Misner et al. \cite{MI70} pp. 704-10, links gravitation by his field equations to the Riemannian metric.
In his approach, the mass distribution in the cosmos into stars, galaxies and their cluster is smoothed out into a mass fluid. He closes the cosmic 3-space into a finite sphere of dimension three, for short the 3-sphere $S^3$. 

The motivation  for considering other spherical cosmic spaces  comes from
observed multipole properties of the cosmic microwave background CMB radiation: Low amplitudes found for the 
lowest multipole orders suggest selection rules from the structure of cosmic 3-space.
If cosmic 3-space is closed on a fraction of the circumference of Einstein's 
3-sphere, any eigenmodes living on it must have shorter intervals of repetition, and must form  a selected subset from all general eigenmodes of the 3-sphere.

All spherical 3-manifolds ${\cal M}$ are variants of Einstein's cosmos and share its positive curvature. They are 
pieces from  the   3-sphere, but are closed by a  homotopic identification of their boundaries. 
Einstein's 3-sphere is simply connected. In contrast,  spherical  topologies
provide  multiple connections and give various fundamental groups \cite{EV04},  \cite{KR10}, see Table 4.

Cosmic topology looks at  CMB radiation on particular manifolds as candidates for cosmic 3-space.  Random CMB amplitudes from the surface of last scattering are simulated with  manifold-specific bases and multipole selection rules. 
To achieve these goals for Platonic manifolds, we present  here new tools needed from topology, group theory and functional analysis. We argue with \cite{KR10}, \cite{KR10A}, \cite{KR10B} that  proper random functions on a spherical 
manifold should be independent of point-symmetric positions. In terms of topology 
this leads from manifolds to orbifolds. We study their novel deck groups in section \ref{Orb}, functional analysis and multipole selection rules in section \ref{Har}, boundary conditions in section \ref{Hom}, and bases
in section \ref{Rec}. In Appendix A we illuminate the concepts of orbifolds and random point symmetry  on the square/torus example. In Appendix B we give the representation for products of Weyl reflections.

\section{Spherical spaces and eigenmodes.}
\label{Spa}
Spherical spaces as topological manifolds are abstractly described as the quotients of the covering  manifold $S^3$ and the group $H$ of order $|H|$ acting fixpoint-free.
The topologies  are then classified as space forms ${\cal M}=S^3/H$. They were characterized with faithful representations of groups $H$ by Wolf in \cite{WO84} pp. 198-230. This characterization leaves open  the geometry of the manifold, and so gives no access to point symmetry. Therefore we prefer a  geometric approach \cite{KR10}, starting from homotopy with Everitt  \cite{EV04}.

Each spherical manifold carries a specific harmonic  basis of eigenmodes, invariant under $H$ and obeying homotopic boundary conditions. 
Harmonic  polynomial bases vanish under the Laplacian on the Euclidean space $E^4$ 
that embeds the 3-sphere.

The eigenmodes of a spherical manifold offer two alternative views  on different domains: (i) On a single closed manifold ${\cal M}$, they  allow to expand square integrable observables. These obey homotopic boundary conditions for functional values on faces and edges.\\ (ii) The covering 3-sphere is 
tiled by copies of ${\cal M}$. Any
eigenmode now must on each tile repeat its value, and fulfill the homotopic 
boundary conditions. It follows that the eigenmodes must display selection rules in comparison to a general polynomial basis on the 3-sphere. This second view allows for  a comparison of observables for different topologies on the same domain, Einstein's 3-sphere.

In topology, these two views present  (i) the local concept of homotopy of ${\cal M}$, and 
(ii) the concept of deck transformations from  the group $H={\rm deck}({\cal M})$ that generate the tiling on the  3-sphere as universal cover. 
Seifert and Threlfall \cite{SE34} pp. 195-8 prove the equivalence of the two views: the fundamental or first homotopy group  $\pi_1({\cal M})$ is isomorphic to the group of deck transformations ${\rm deck}({\cal M})$ that generates the tiling on the cover.

\section{Spherical Platonic 3-manifolds.}
\label{Spe}

We study  here the family of Platonic spherical 3-manifolds.
For each Platonic polyhedron we construct in \cite{KR10} on the 
universal cover, the 3-sphere, a unique group $H$  of fixpoint-free deck transformations 
acting on the 3-sphere 
as a subgroup of a Coxeter group $\Gamma$. 
By the theorem from \cite{SE34}, the deck groups $H$ are isomorphic to, and were derived from,  the fundamental or first homotopy groups constructed by Everitt \cite{EV04}.

Coxeter groups $\Gamma$ \cite{HU90} will become a main tool of the following  analysis.
They  act  on Euclidean 4-space with coordinates $x=(x_0, x_1, x_2, x_3)$, and on the 3-sphere, by four involutive Weyl reflections $(W_1, W_2, W_3, W_4)$ in hyperplanes, see Table \ref{table:Table4}.
The Coxeter diagram encodes the relations between the Weyl reflections, associated to  its four nodes.
A connecting line of two nodes implies $(W_iW_{i+1})^3=e$, 
a connecting line with integer superscript  $k > 3$ implies $(W_iW_{i+1})^k=e$,
Weyl reflections for nodes not connected by lines commute with one another.
Table \ref{table:Table4} reviews the relation of the Platonic polyhedra to Coxeter groups.
In Table \ref{table:Table5}  we list the four unit vectors $a_j \in E^4$, perpendicular to the Weyl 
reflection hyperplanes, for each Coxeter group.
The Coxeter group tiles the 3-sphere into $|\Gamma|$ Coxeter simplices. The initial polyhedron consists of
those Coxeter simplices which share a vertex at $x=(1,0,0,0)$.  
In topology we prefer  orientable manifolds. In the Coxeter groups this means that
we restrict attention to the subgroups generated by an even number of Weyl reflections. We call the
corresponding subgroups $S\Gamma$, where $S$ stands for unimodularity of the defining representation 
on the Euclidean space $E^4$. The order of these subgroups is $|S\Gamma|=|\Gamma|/2$. 
A set of  generators of all unimodular Coxeter groups with four Weyl reflections is given by 
\begin{equation}
\label{or1}
 S\Gamma: \{ (W_1W_2), (W_2W_3), (W_3W_4) \}.
\end{equation}
Note that any product $(W_iW_j)$ leaves invariant the intersection of the two Weyl reflection hyperplanes 
perpendicular to the vectors $\{a_i, a_j\}$. The representation of the product is given in Appendix B,

\begin{figure}[tbp]
\begin{center}
\includegraphics[width=0.7\textwidth]{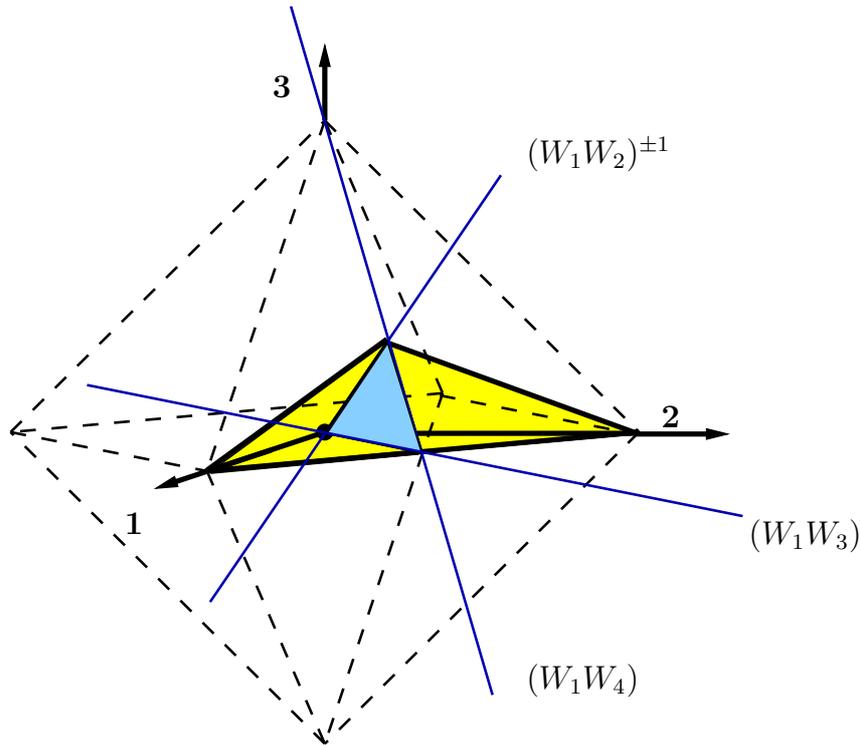} 
\end{center}
\caption{\label{fig:Fig3}
{\bf Octahedral orbifold}. Face gluing deck rotations of the octahedral duplex
orbifold (yellow) $N10$ in Euclidean version. The corresponding  four products $ (W_1W_3), 
(W_1W_2)^{\pm 1}, (W_1W_4)$ of two Weyl reflections from $\Gamma =\circ - \circ  \stackrel{4}{-}\circ -\circ$ are marked 
by rotation axes, intersection lines of their Weyl planes. The three intersections of these lines with the orbifold
form the  glue  triangle (light blue area). Its inner edge points have the orders (2, 3, 2) 
of their covering rotations.}
\end{figure}

\begin{figure}[tbp]
\begin{center}
\includegraphics[width=0.6\textwidth]{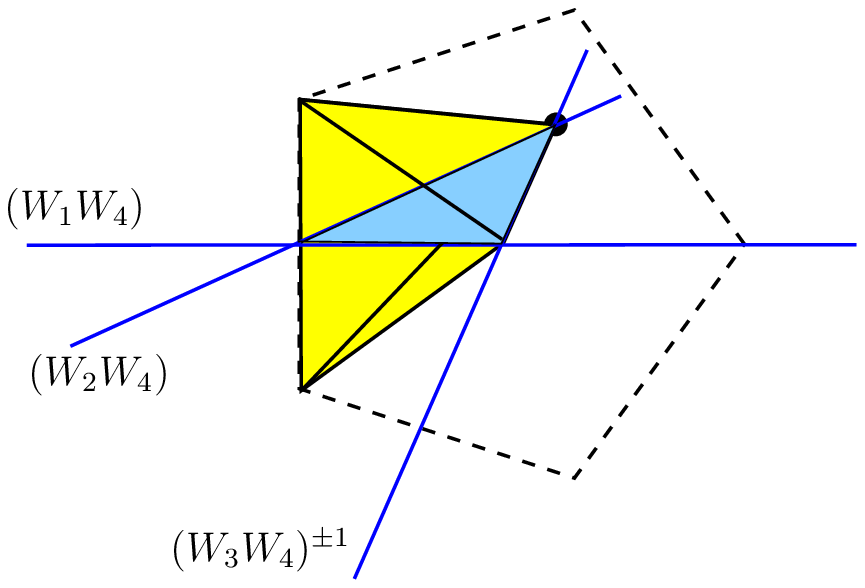} 
\end{center}
\caption{\label{fig:Fig4}
{\bf Dodecahedral orbifold}. Face gluing deck rotations of the dodecahedral duplex orbifold (yellow) $N11$, attached to a 
dodecahedral face, in Euclidean version. The corresponding  four covering products $ (W_2W_4), 
(W_3W_4)^{\pm 1}, (W_1W_4)$ of two Weyl reflections from 
$\Gamma =\circ - \circ -\circ \stackrel{5}{-}\circ $ are marked 
by the intersection lines of their Weyl planes. The three intersections of these lines with the orbifold
form the glue triangle (light blue area). Its inner edge points have the orders (2, 5, 2) 
of their covering rotations.}
\end{figure}

\section{Duplices under the point group tile a polyhedron.}
\label{Dup}

In \cite{KR10} we in addition introduce for random functions  the notion of random symmetry under 
the point group $M$ of these manifolds. We argue there that the values of a proper random function on a polyhedral manifold with point symmetry should be independent of operations from $M$. Therefore one should explore $M$ for the analysis of the CMB radiation. The point groups are the tetrahedral group $A(4)$ for the tetrahedral 3-manifold, the cubic group $O$ 
for the cubic and octahedral 3-manifolds, and the icosahedral group ${\cal J}$ for the spherical 
dodecahedron. These point groups are unimodular subgroups w.r.t the 2-sphere 
\begin{equation}
\label{or1a}
x_1^2+x_2^2+x_3^2=1.
 \end{equation}
and may be characterized on 3-space by unimodularity $S$ and Coxeter subdiagrams as
\begin{equation}
\label{or1b}
A(4)=S(\circ-\circ-\circ),\: O=S(\circ\stackrel{4}{-}\circ-\circ)\sim S(\circ-\circ\stackrel{4}{-}\circ),
\:{\cal J}=S(\circ-\circ\stackrel{5}{-}\circ).
\end{equation}

The  action of the point group $M$ on the  Platonic proto-polyhedron can now be decomposed into a fundamental domain and its images under $M$. The shape of the fundamental domain is not unique, but we can choose it  in compact, convex and polyhedral form. Each fundamental domain for $M$  we take as a duplex inside the proto-polyhedron, formed by gluing a Coxeter simplex and its
mirror image  under reflection in one simplex face.  

{\bf Prop 1: Fundamental domains under point groups}:
For each Platonic polyhedron, a duplex fundamental domain of their point group $M$ may be chosen as shown 
in Figures \ref{fig:Fig1} - \ref{fig:Fig4}.  All of them are again simplices  with four faces.
Images under $M$ of this  duplex tile the polyhedron. 

\section{Orbifolds under $S\Gamma$ tile the 3-sphere.}
\label{Orb}

The Platonic 3-manifolds under their deck group $H$  in turn tile the 3-sphere into $|H|$ copies of the Platonic polyhedron as prototile. 
These tilings into Platonic polyhedra are the m-cells, $m=|H|$, described in \cite{SO58} and \cite{KR10}.
By composing the polyhedra, tiled into duplices, as tiles of the $|H|$-cell  tilings, it follows that the 3-sphere is tiled into $|S\Gamma|=|M| \cdot |H|$ duplices. 
With respect to actions on the 3-sphere, the Coxeter duplices, which originated as fundamental domains of the point group acting on the Platonic prototile, now become  the 
fundamental domains of the much bigger groups $S\Gamma$. We shall identify them as orbifolds.

\subsection{The action of deck and point groups.}
The subgroup $H < S\Gamma$  of deck transformations for a given 3-manifold  acts on the 3-sphere without fixpoints.
The point group $M < S\Gamma$  by definition preserves the center of the polyhedron.  It  follows \cite{KR10B} that the intersection 
of these two subgroups consists of the identity,  $H \cap M=e$. 

Any image of the orbifold as proto-duplex under $S\Gamma$ on the 3-sphere has a unique compound address, composed of a unique point group element $p$ acting on the initial polyhedron, followed by a  unique deck transformation $h$ from  the initial polyhedron to an image on the $|H|$-cell tiling. This leads to the following general conclusion on the group structure:

{\bf Prop 2: Unimodular Coxeter groups are products of subgroups }: Given a prototile duplex, its image tile in the Coxeter duplex tiling  of the 3-sphere results from the action of a  group element from $S\Gamma$, uniquely factorized into a deck and a point transformation from $H \cdot M$.  This means that the 
elements  $g \in S\Gamma$ obey the unique subgroup product law 

\begin{equation}
\label{or1bb}
g\in S\Gamma:\: g= h  p,\:  h \in H,\: p \in M,\: H \cap M=e,\: |S\Gamma|=|H|\cdot  |M|.
\end{equation}
{\bf Proof}: The group $S\Gamma$ with subgroups $H,\: M$ all have faithful (one to one)  representations, \cite{WO84} p. 138, on $E^4$.
This implies  that any image under $g \in S\Gamma$ of the orbifold as proto-duplex has a unique 
compound address as a product $g=h p,\:  h\in H, p\in M$. Similarly one can construct for the same image 
a unique compound address $g=p' h'$.

This  product structure $S\Gamma=H \cdot M$ differs from   a direct or semidirect product. The two subgroups $H, M$ do not commute, and  the group $H$ is not invariant under conjugation with elements from $M$,  except in  
case of the cubic Coxeter group, analyzed in \cite{KR10}. Eq. \ref{or1bb} shows that each of the subgroups generates the cosets for its partner.

\subsection{Topology and deck groups for orbifolds.}
\label{Top}

We turn to the topological significance of the fundamental duplex domains for the group $S\Gamma$.
A duplex of a unimodular Coxeter group on the 3-sphere 
cannot  form a topological 3-manifold, because under the point group action it exhibits fixpoints of finite order $k>1$ on its boundaries. The order $k$ of a point is defined as the order of its stabilizer.
Points of order $k=1$ in topology are called regular, for order $k>1$ singular, \cite{RA94} pp. 664-6.
We propose here to move in topology from the standard space forms to  orbifolds. 
This concept is illustrated on the square and torus in Appendix A.

We refer to Montesinos \cite{MO87} pp. 78-97 and 
to Ratcliffe \cite{RA94} pp. 652-714 for the introduction and mathematical terms associated with  orbifolds. The notion  includes a manifold structure and a covering, but  
admits singular points of finite order.

We claim: 

{\bf Prop 3: Spherical Platonic 3-manifolds with point symmetry  are appropriately described  by orbifolds.} 

We  shall  demonstrate orbifold coverings by deck transformations from $S\Gamma$, and derive 
the  harmonic analysis for their use in cosmic topology.

In figures \ref{fig:Fig1},  \ref{fig:Fig2}, \ref{fig:Fig3}, \ref{fig:Fig4} we reproduce, with minor changes, based on \cite{KR10A} a set of fundamental duplex domains for the point groups. Now we interprete them as 3-orbifolds under  $S\Gamma$ acting on the 3-sphere, based on  the tetrahedron, the cube, the octahedron or the dodecahedron,  drawn  in their Euclidean version. The edges of the Platonic polyhedra are given in dashed lines, the orbifolds are marked by  yellow color. 

For topological 3-manifolds, the deck group on the 3-sphere is generated by the operations which map the pre-image
of the manifold to all its face neighbours. We present  a corresponding analysis for the deck group of an
orbifold on the 3-sphere. 

We use the tetrahedral manifold $N2$ from Table \ref{table:Table4}  for demonstration of the covering.
Here the Coxeter group is $S(5)$, the symmetric group on $5$ objects, with unimodular subgroup $S\Gamma=A(5)$, the group of even permutations. In Fig. \ref{fig:Fig1} we mark the four vertices of the tetrahedron by the numbers 
$1,2,3,4$. A different  enumeration is used in  \cite{KR10}, Fig. 4.
The Weyl reflection operators are in one-to-one correspondence to the transpositions
\begin{equation}
 \label{or1c}
W_1=(1,2),\: W_2=(2,3),\: W_3=(3,4),\: W_4=(4,5).
\end{equation}

From eq. \ref{or1b} we find for the four covering deck rotations of this orbifold the simple expressions
\begin{eqnarray}
 \label{or1d}
&&(W_1W_3)=(1,2)(3,4),
\\ \nonumber 
&&(W_1W_2)=(1,2)(2,3)=(1,2,3),\: (W_1W_2)^{-1}=(2,3)(1,2)=(3,2,1),
\\ \nonumber 
&&(W_1W_4)=(1,2)(4,5),
\end{eqnarray}
compare Fig. \ref{fig:Fig1}. These even permutations generate $A(5)$.

We return to all four orbifolds associated with Platonic polyhedra and  denote them by $N8, N9$, $N10$ as in \cite{KR10A} and by $N11$ for 
the Platonic dodecahedron.
Each face of the orbifold as proto-tile of the duplex tiling is covered by a face-neighbour, a  copy of the orbifold. 
Instead of drawing the neighbours we give in the figures the deck rotations from $S\Gamma$ which map the 
preimage into its face neighbours. These deck rotations have axes which intersect with faces of the orbifold. In the figures we denote the rotation axes 
by blue lines and give in each case the even product of Weyl reflections. The first three deck rotations 
all have axes passing through the center of the polyhedron and generate  the point group $M$.
The last even product always involves a Weyl reflection that passes through an outer face of the Platonic polyhedron. The  rotation  containing this Weyl reflection
has order $k=2$. It transforms the orbifold into a face neighbour  inside  a  new Platonic neighbour polyhedron.   By examining the deck operations and comparing with eq. \ref{or1}, it can be verified 
in each case that the four deck operations that cover the faces of the orbifold generate the full unimodular Coxeter group $S\Gamma$. This is in full analogy to the role of  deck groups $H$  for spherical 
3-manifolds \cite{KR10}.

The intersections of the covering  rotation axes with the 3-orbifold determine singular  points and their order. The center point of the initial spherical proto-polyhedron, chosen as $x=(1,0,0,0)$,  has the maximal order $k=|M|$. For the orbifolds under inspection, more  fixpoints appear on the inner  points of edges of the glue  triangle of each duplex, with area 
marked in the figures in light blue. The order $k$ of these fixpoints agrees with the order of the covering rotation. Note that, in contrast to the Euclidean drawings, we always infer the order $k$ of the rotations 
from the Coxeter group relations and their spherical settings on the 3-sphere.
We give the covering rotations and orders of these  fixpoints in 
Table \ref{table:Table1}.

The obvious rotation axes from the point group $M$ determine additional  singular points 
on inner positions of edges of the orbifolds. For all  four orbifolds we find:

\begin{table}[t]
$
\begin{tabular}{|l|l|l|l|l|l|}\hline
Coxeter & $|S\Gamma|$= & polyhedron, & $M, |M|$ & deck generators&order k\\ 
diagram $\Gamma$&$|H|\cdot |M|$&orbifold &&of orbifold&
\\ \hline 
$\circ -\circ -\circ - \circ$               & $5 \cdot 12$  & tetrah., N8 & $A(4), 12$ & $(W_1W_3), (W_1W_2)^{\pm 1}, (W_1W_4)$
&(2, 3, 2)
\\  \hline
$\circ \stackrel{4}{-} \circ -\circ -\circ$ & $8 \cdot 24$  & cube, N9 & $O, 24$ & $(W_1W_3), (W_1W_2)^{\pm 1}, (W_1W_4)$ 
&(2, 4, 2) 
\\  \hline
$\circ -\circ \stackrel{4}{-}\circ - \circ$ & $24 \cdot 24$ & octah., N10 &$O$, 24 & $(W_1W_3), (W_1W_2)^{\pm 1},  (W_1W_4)$
&(2, 3, 2)\\ \hline
$\circ -\circ - \circ\stackrel{5}{-}\circ $ & $120 \cdot 60$ & dodecah., N11 &${\cal J}$, 60 
& $(W_2W_4), (W_3W_4)^{\pm 1},  (W_1W_4)$&(2, 5, 2)
\\ \hline
\end{tabular}
$
\caption{\label{table:Table1}
4 Coxeter groups $\Gamma$, $|S\Gamma|$, Platonic polyhedra, orbifolds, point groups $M$, 
deck generators of orbifolds, and selected  orders $k$ for inner points on edges of the glue triangle of four  duplex orbifolds $N8, N9, N10, N11$. $A(4) $ is the tetrahedral, $O$ the cubic, ${\cal J}$ the icosahedral rotation group. The center of the polyhedron has order $|M|$, the inner points of the faces of the orbifolds have order $k=2$, inner points of the orbifolds are regular. Note that the Weyl reflections $W_l$ depend on the Coxeter group chosen, see Table \ref{table:Table5}.}
\end{table}

{\bf Prop 4: Deck transformations of orbifolds generate unimodular Coxeter groups}: Deck rotations  cover  the four faces of the four Platonic 3-orbifolds, they generate the duplex tiling of $S^3$ by all elements of the corresponding unimodular Coxeter deck groups $S\Gamma$.

It follows that the orbifolds can abstractly be characterized as quotient spaces $S^3/S\Gamma$.

\section{Harmonic analysis on orbifolds.}
\label{Har}

The harmonic analysis can follow the two views described in section \ref{Spa}.
Clearly the homotopic boundary conditions imply  selection rules compared to  the full basis on the 3-sphere.
The  CMB radiation on an assumed  topology usually is  modelled by random coefficients in the specific  basis.
Here we analyze the basis construction, but postpone any numerical modelling.

On the 3-sphere, after replacing the coordinates $x$  by  matrix coordinates $u=u(x) \in SU(2,C)$ \cite{KR10}, see Appendix B, the harmonic basis may be spanned by Wigner polynomials $ D^j_{m_1m_2}(u)$ \cite{ED57} or by spherical
polynomials $ \psi (jlm)(u)$, with linear relations 

\begin{eqnarray}
 \label{A0}
&&\psi (jlm)(u)= \sum_{m_1m_2} D^j_{m_1m_2}(u) \langle j-m_1jm_2|lm\rangle (-1)^{(j-m_1)},
\\ \nonumber
&& D^j_{m_1m_2}(u)= \sum_{lm} \psi (jlm)(u)\langle j-m_1jm_2|lm \rangle (-1)^{(j-m_1)},\: 0\leq l\leq 2j,
\end{eqnarray}
and with summations restricted by the Wigner coefficients \cite{ED57}.
The spherical polynomials transform under rotations of 3-space $(x_1, x_2, x_3)$ like the spherical harmonics $Y^l_m$ \cite{KR10}, used in the CMB data analysis. Point symmetry selects 
the lowest multipole order, compare \cite{KR10} Table 3, \cite{KR10A}.
 The Wigner polynomial basis is suited for the projection
of the subbasis invariant under the deck group $H$.
In the cubic case, the deck group 
of the 3-manifold $N3$ is the quaternion group $H=Q$. Invariant under conjugation with  the cubic point group $O$, it forms a semidirect product $S\Gamma=(Q \times_s O)$.

The harmonic analysis on the four spherical orbifolds $N8, N9, N10, N11$ is given by polynomials invariant under 
their unimodular Coxeter group $S\Gamma$. These and only these polynomials repeat their values on 
any Coxeter duplex from the tiling.

{\bf Prop 5: Harmonic analysis on spherical 3-orbifolds}: The harmonic analysis on a spherical orbifold 
is spanned by harmonic polynomials, invariant under its group of deck transformations $S\Gamma$.

We have seen in Prop. 2 that any element of $S\Gamma$ admits a unique factorization. For the projectors 
to the identity representation of $S\Gamma$ we claim by use of eq. \ref{or1a}:

{\bf Prop 6: Factorization of projectors}:  The projector to the identity representation, denoted by $\Gamma_1$, for the unimodular Coxeter group  $S\Gamma$ factorizes into the product of the projectors to the identity representations $\Gamma_1$ for the two subgroups $H, M$, 
\begin{equation}
\label{or2}
P^{\Gamma_1}_{S\Gamma}= P^{\Gamma_1}_{H} P^{\Gamma_1}_{M}=P^{\Gamma_1}_{M} P^{\Gamma_1}_{H}.
\end{equation}

{\em Proof}: In the group operator algebra of the unimodular Coxeter group $S\Gamma$ we have 
from eq. \ref{or1b} the factorization
\begin{eqnarray}
 \label{or3}
&& P^{\Gamma_1}_{S\Gamma}=\sum_g T_g= \sum_{h,\: p} T_{h p}= \sum_{p,\: h} T_{p h},\:\: h\in H, p \in M,
\\ \nonumber 
 && =\sum_{h,\: p} T_h T_p= (\sum_h T_h)(\sum_p T_p)=(\sum_p T_p)(\sum_h T_h)=P^{\Gamma_1}_{H}\: P^{\Gamma_1}_{M} =P^{\Gamma_1}_{M} P^{\Gamma_1}_{H}.
\end{eqnarray}
This result greatly simplifies the projection to the identity representation:

The projectors $P^{\Gamma_0}_{H}$  for the group H of deck transformations are given in  \cite{KR10} in the Wigner polynomial basis, see eq. \ref{A3}.
So it remains to pass with eq. \ref{A0} from the Wigner  to the spherical basis, and then to apply the projectors $P^{\Gamma_0}_{M}$ of the point group.

{\bf Prop 7: Orbifolds give sharp multipole selection rules}: $S\Gamma$-invariant polynomials by their point group $M$-invariance select  a lowest non-zero multipole order $l$, see \cite{KR10} Table 3. 
This projection  is carried out  in \cite{KR10A} for the spherical cubic 3-orbifold and multipole order 
$0\leq l\leq 8$. The results are reproduced here in Tables \ref{table:Table2} and \ref{table:Table3}  in explicit multipole order.
The analysis   can be extended by the method described in section \ref{Rec}.

A similar analysis applies to the spherical tetrahedral, octahedral and dodecahedral 3-orbifolds. 

\begin{table}[t]
$
 \begin{array}{l|l}
l&Y^{\Gamma_1,l}=\sum_m a_{lm}Y^l_m(\theta, \phi)\\ \hline
0&Y^0_0\\
4&\sqrt{\frac{7}{12}}Y^4_0+\sqrt{\frac{5}{24}}(Y^4_4+Y^4_{-4})\\
6&\sqrt{\frac{1}{72}}Y^6_0-\sqrt{\frac{7}{144}}(Y^6_4+Y^6_{-4})\\
8&\frac{1}{64}\sqrt{33} Y^8_0+\frac{1}{12}\sqrt{\frac{21}{2}}(Y^8_4+Y^8_{-4})+
\frac{1}{24}\sqrt{\frac{195}{2}}(Y^8_8+Y^8_{-8})\\
\end{array}
$

\caption{\label{table:Table2} The lowest cubic $O$-invariant  spherical harmonics $Y^{\Gamma_1,l}$, expressed by spherical harmonics $Y^l_m$.}
\end{table}

\begin{table}[t]
$
 \begin{array}{l|l|l}
2j&l&\psi^{0,\Gamma_1,2j}=\sum_{l} b_{l} R_{2j+1\; l}(\chi)Y^{\Gamma_1,l}(\theta, \phi)\\ \hline
0&0& R_{1 0} Y^{\Gamma_1, 0}\\
4&0,4&\sqrt{\frac{2}{5}}R_{5 0}Y^{\Gamma_1, 0}+\sqrt{\frac{3}{5}}R_{54}Y^{\Gamma_1, 4}\\
6&0,4, 6&\sqrt{\frac{1}{7}}R_{7 0}Y^{\Gamma_1, 0}-
\sqrt{\frac{6}{11}}R_{7 4}Y^{\Gamma_1, 4}-\sqrt{\frac{24}{77}}R_{76}Y^{\Gamma_1, 6}\\
8&0,4, 6, 8&\frac{4}{3}\sqrt{\frac{1}{110}}R_{9 0}Y^{\Gamma_1, 0}
-\frac{12}{11}\sqrt{\frac{3}{65}}R_{9 4}Y^{\Gamma_1, 4}\\
&&+\frac{8\cdot 19}{165}R_{96}Y^{\Gamma_1, 6}
+\frac{4}{5}\sqrt{\frac{1}{33\cdot 13}}R_{98}Y^{\Gamma_1, 8}\\
 \end{array}
$
\caption{\label{table:Table3} The lowest $(S\Gamma=(Q \times_s O))$-invariant polynomials $\psi^{0,\Gamma_1,2j}$
of degree $2j$  on the 3-sphere, expressed by  the cubic invariant  spherical harmonics 
from  Table \ref{table:Table2}. $(Q \times_s O)$-invariance enforces superpositions of several cubic invariant  spherical harmonics.}
\end{table}

\section{Homotopic boundary conditions from orbifolds.}
\label{Hom}

From \cite{KR10} we know that a fixed geometric shape of a Platonic 3-manifold can have different and inequivalent topologies, characterized by different groups of homotopies and deck transformations $H$. 
These differences give rise to different homotopic boundary conditions. We now examine the boundary conditions for orbifolds.

We have seen that the orbifold  is covered face-to-face by rotational  images.
It follows, as in the case of 3-manifolds, that the topology on 3-manifolds implies homotopic boundary conditions on the faces of the 3-orbifolds. Since with the orbifolds  we introduce point symmetry in addition to 
deck transformations, we find from the arguments given in \cite{KR10} for point symmetry,

{\bf Prop 8: Topological universality from point symmetry}: If we demand, for  a function on a given Platonic 3-manifold, point symmetry under $M$ in addition to the boundary conditions set by homotopy on faces and edges, 
then new boundary conditions apply universally, that is, independent of the specific group of deck transformations chosen. 

{\bf Prop 9: Universal homotopic boundary conditions from 3-orbifolds}: If in addition to homotopy we demand on the manifold symmetry under the  rotational point group $M$, the homotopic boundary conditions for different deck 
and homotopy groups on the same Platonic geometrical shape coincide with one another and reduce to  the homotopic boundary conditions for the orbifolds. Their boundary conditions are determined by the 
generators in Table \ref{table:Table1} of the covering rotations.

\begin{table}[t]
$
\begin{tabular}{|l|l|l|l|l|l|}\hline
Coxeter diagram $\Gamma$ & $|\Gamma|$ & polyhedron ${\cal M}$ & $H={\rm deck}({\cal M})$ & $|H|$ & Reference \\ \hline
$\circ -\circ -\circ - \circ$               & $120$  & tetrahedron $N1$ & $C_5$         & $5$ & \cite{KR08} \\  \hline
$\circ \stackrel{4}{-} \circ -\circ -\circ$ & $384$  & cube $N2$        & $C_8$         & $8$ & \cite{KR09} \\                                                           &        & cube $N3$        & $Q$           & $8$ &\cite{KR10} \\  \hline
$\circ -\circ \stackrel{4}{-}\circ - \circ$ & $1152$ & octahedron $N4$  & $C_3\times Q$ & $24$ &\cite{KR10} \\
                                            &        & octahedron $N5$  & $B$           & $24$ &\cite{KR10} \\
                                            &        & octahedron $N6$  & ${\cal T}^*$  & $24$&\cite{KR10} \\ \hline
$\circ -\circ -\circ \stackrel{5}{-} \circ$  & $120^2$  & dodecahedron$N1'$& $ {\cal J}^*$ & $120$&\cite{KR05}\\
\hline
\end{tabular}
$
\caption{\label{table:Table4}
4 Coxeter groups $\Gamma$, 4 Platonic polyhedra ${\cal M}$, 7 deck groups $H={\rm deck}({\cal M})$ of order $|H|$.
$C_n$ denotes a cyclic, $Q$ the quaternion, ${\cal T}^*$ the binary tetrahedral, ${\cal J}^*$ the binary icosahedral, S$\Gamma$ a unimodular Coxeter group. The symbols $Ni$ are taken  from  \cite{EV04}.}
\end{table}

\begin{table}
\begin{equation*}
\begin{array}{|l|l|l|l|l|} \hline
\Gamma & a_1 & a_2 &a_3& a_4\\ \hline
\circ -\circ -\circ - \circ &  (0,0,0,1)&(0,0,\sqrt{\frac{3}{4}},\frac{1}{2})
& (0,\sqrt{\frac{2}{3}},\sqrt{\frac{1}{3}},0)& (\sqrt{\frac{5}{8}},\sqrt{\frac{3}{8}},0,0)\\ \hline
\circ \stackrel{4}{-} \circ -\circ -\circ&  (0,0,0,1)& (0,0,-\sqrt{\frac{1}{2}},\sqrt{\frac{1}{2}})
&(0,\sqrt{\frac{1}{2}},-\sqrt{\frac{1}{2}},0)&(-\sqrt{\frac{1}{2}},\sqrt{\frac{1}{2}},0,0)\\ \hline
\circ -\circ \stackrel{4}{-}\circ - \circ& (0,\sqrt{\frac{1}{2}},-\sqrt{\frac{1}{2}},0)&(0,0,-\sqrt{\frac{1}{2}},\sqrt{\frac{1}{2}})
&(0,0,0,1)& (\frac{1}{2},\frac{1}{2},\frac{1}{2},\frac{1}{2})\\ \hline
\circ -\circ -\circ \stackrel{5}{-} \circ& (0,0,1,0)&(0,-\frac{\sqrt{-\tau+3}}{2},\frac{\tau}{2},0)
&(0,-\sqrt{\frac{\tau+2}{5}},0,-\sqrt{\frac{-\tau+3}{5}})&(\frac{\sqrt{2-\tau}}{2},0,0,-\frac{\sqrt{\tau+2}}{2}) \\ \hline
\end{array}
\end{equation*}
\caption{\label{table:Table5}
The Weyl vectors $a_s$  
for the 4 Coxeter groups $\Gamma$ from Table \ref{table:Table4}  with  $\tau:=\frac{1+\sqrt{5}}{2}$.
}
\end{table}

\section{Recursive computation of the bases invariant under the orbifold deck groups $S\Gamma=H\cdot M$.}
\label{Rec}

We describe the recursive construction of $S\Gamma$-invariant bases by use of the factorization
given in Prop. 4. This recursive construction was used to obtain the results of Tables 2, 3. The multiplicity $m(l,\Gamma_1)$ of the identity representation $\Gamma_1$ of $M$ for given multipole order $l$ is given 
in \cite{LA74} pp. 436-8.

Our basic tool are the relations \cite{KR10} between Wigner  and spherical harmonic polynomials on the 3-sphere given in eq. \ref{A0}.

(i) We start from a linear combination of spherical harmonics of multipole order $l$,  invariant under the point group $M$, and construct with its coefficients $a_{lm}$  a $M$-invariant linear combination of spherical polynomials $\psi_{j l m}(u),\: l\leq 2j$ eq. \ref{A0},  on the 3-sphere \cite{KR10},
\begin{equation}
\label{A1}
\psi_{j,l,\Gamma_1}(u)= \sum_m' a^{\Gamma_1}_{lm'}\psi_{j,l,m'}(u)
\end{equation}
The starting linear combinations of spherical harmonics  for lowest multipole order $l$ for the point groups involved can be found in  the literature.
Note that the coefficients $a^{\Gamma_1}_{lm'}$ are independent of the degree $2j$.

(ii) Next we transform with eq. \ref{A0} from the spherical  to the Wigner basis and apply the projector $P^{\Gamma_1}_H$ 
from \cite{KR10} for $H$-invariance. If the resulting  function does not vanish, we 
transform it back to the spherical basis. In this way we find from the starting function a new one,  invariant from Prop. 4 under both $M$ and $H$ and hence under $S\Gamma$, given by 
\begin{eqnarray}
\label{A2}
&&\psi_{j,\Gamma_1}(u)= \sum_{i} \left[\sum_m a^{\Gamma_1}_{l+i, m} \psi_{j,l+i,m}(u)\right], 
\\ \nonumber
&&a^{\Gamma_1}_{l+i, m}
= \sum_{m',m_1',m_2',m_1,m_2}
a_{lm'}  \langle j-m_1'jm_2'|lm'\rangle (-1)^{(j-m_1')}
\\ \nonumber
&&\langle jm_1'm_2'|P^{\Gamma_1}_{H}|j m_1m_2\rangle
\langle j-m_1jm_2|l+i\: m\rangle (-1)^{(j-m_1)},
\end{eqnarray}
The polynomial eq. \ref{A2} if non-vanishing can be normalized.

The Wigner coefficients are given in \cite{ED57}, and the matrix elements of the projector $P^{\Gamma_1}_H$ for the identity representation of the group $H$ are given in the Wigner polynomial basis by
\begin{equation}
\label{A3}
\langle jm_1'm_2'|P^{\Gamma_1}_{H}|j m_1m_2\rangle
=\frac{1}{|H|} \sum_{h =(h_l,h_r)\in H} D^j_{m_1'm_1}(h_l^{-1})D^j_{m_2'm_2}(h_r). 
\end{equation}
and specified in \cite{KR10} for  each Platonic  3-manifold. Due to universality Prop. 5, we can choose the most convenient deck group $H$ for a fixed geometric shape.
The recursion relation eq. \ref{A2} involves  Wigner coefficients, the elements of the group $H$  given as pairs
$h=(h_l, h_r)$ in \cite{KR10}, and Wigner $D^j$-representations for the group $SU(2,C)$. 
The $S\Gamma$-invariant bases appear as linear combinations of $M$-invariant spherical functions
with fixed multipole order $l+i$.

(iii) Moreover, since the point group action cannot change the multipole order,
each new partial sum  of eq. \ref{A2} in square brackets for fixed $l+i$ must separately  be invariant under the point group $M$. This allows to restart the computation with $l \rightarrow l+i$ by going again from spherical to Wigner polynomials, followed by projection of an invariant under $H$. In this way we can increase the polynomial degree $2j$.
By character technique we can control the number of invariants for given degree $2j$ of the polynomials.

\section{Conclusion.}
\label{Con}

We propose the  interpretation and use of 3-orbifolds in cosmic topology along two lines:

(I) The notion of  a topological 3-manifolds with (random) point symmetry $M$ is reformulated in terms   of topological 
3-orbifolds. The relevant group of deck transformations is a unimodular Coxeter group $S\Gamma$.
This group generates, from a Coxeter duplex orbifold as prototile, a tiling of the 3-sphere into $|S\Gamma|$ copies.
The group $S\Gamma=H \cdot M$ combines  the group $H$ of polyhedral deck transformations with the point group $M$
of the polyhedron. The Platonic  3-manifolds under the assumption of random point symmetry shrink into 3-orbifolds, and
their harmonic bases live on a fraction $1/|S\Gamma|$  of the 3-sphere, see Table 1. They are more selective than those for the 3-manifolds, and therefore easier to test.

For the deck group $S\Gamma= H \cdot M$  we construct a new  harmonic analysis which can model the CMB radiation.
Point symmetry implies sharp multipole selection rules, and topological universality: all fundamental groups
for the same geometrical polyhedral shape produce the same boundary conditions.
There is only 
a single point-symmetric harmonic basis, invariant under both $H$ and $M$. This result is demonstrated  in Tables \ref{table:Table2}, \ref{table:Table3} from  \cite{KR10A} for orbifolds from spherical cubes.

(II) We can retain  the strict original notion of topological 3-manifolds. The new  harmonic basis, 
characterized by point symmetry, forms a   subbasis of the harmonic analysis. If it fails to model the CMB fluctuations, 
one can augment it by the larger  basis of \cite{KR10} for the 3-manifold without point symmetry.

\newpage

\begin{figure}[tb]
\begin{center}
\includegraphics[width=0.4\textwidth]{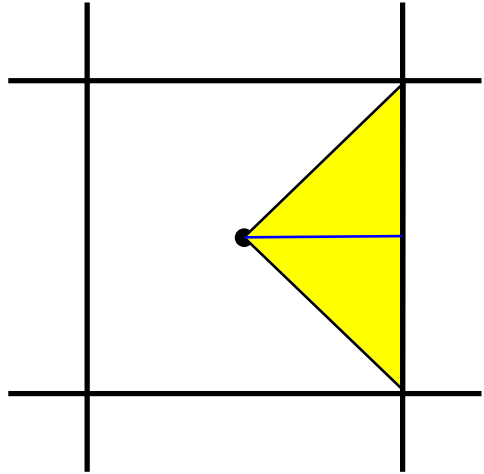} 
\end{center}
\caption{\label{fig:Fig5}
{\bf From the square to its orbifold}. The square in the Euclidean plane. Parallel edges are identified by homotopy to yield the 2-torus $T^2$.
A $(\pi/2)$-sector (yellow) forms a fundamental domain under the group $C_4$ of 4fold rotations.
In topology this sector is an orbifold. The center point under $C_4$ has order $4$.}
\end{figure}

\section{Appendix A: From square and torus to orbifold.}
\label{AppA}

{\bf From square to torus}: The square on the Euclidean plane $E^2$, see figure \ref{fig:Fig5},  is closed into a topological manifold of curvature zero as follows: 
We identify the two pairs of parallel edges - this gluing generates  homotopy. The resulting topological 
manifold is the 2-torus $T^2$, it is finite but unbounded. The 2-torus on its surface admits two types of closed loops,
whose multiple windings are generated by two commuting infinite cyclic groups $C_{\infty} \times C_{\infty}$.
This is the homotopy group of the 2-torus, $\pi_1(T^2)=C_{\infty} \times C_{\infty}$. 
The 2-torus when unfolded back into its universal cover, the plane $E^2$, becomes a prototile of a square tiling. Its repetition pattern consists of two 
infinite translation groups in perpendicular directions. The  group that generates the square tiling
is the discrete two-fold translation or deck group, it is again ${\rm deck}( T^2)=C_{\infty} \times C_{\infty}$. The group of homotopic windings of the 2-torus and the deck group are isomorphic and illustrate the general theorem of Seifert and Threlfall \cite{SE34} pp. 195-8.

{\bf Fourier basis}: Next turn to functions on the square and on the 2-torus. Complex-valued  functions have the exponential basis of the twofold periodic Fourier series. The basis  
obeys homotopic boundary conditions: it repeats its values on parallel edges
of the square tiling. This property extends to any function that can be written as a Fourier series.
 
{\bf Point symmetry}: Now we note that the square has a point symmetry: Multiples of the rotation 
$R(\phi),\:\phi= \pi/2$ map the square into itself while keeping its center. The relevant point group is the cyclic group $C_4$.
If we inscribe into the square a $(\pi/2)$-sector (yellow) by connecting two endpoints of an edge to the center,
we can reach any other points of the square from this sector by applying the four rotations from $C_4$. The sector in topology 
is called an orbifold.

{\bf Random functions}: Turn in particular to random functions $f^{\rm random}(x)$ on the square. If we want to attach them to the 2-torus as 
topological manifold, we must demand for functional values of $f^{\rm random}(x)$ the homotopic boundary conditions
of twofold periodicity,
equivalent to allowing for an  expansion into a twofold Fourier series. The Fourier coefficients may be chosen at random. 
 
{\bf Random point symmetry}: Imagine a random function $f^{\rm random}(x)$ on the square and apply to it the rotation $R(\pi/2)$. 
As edges of the square are mapped into edges, the rotated function $\tilde{f}^{\rm random}(x):=f^{\rm random}(R^{-1}(\pi/2)x)$  is another random function on the same square. To render  a proper random function
independent of this point rotation, we must add the values of both functions, and, extending the argument to all four rotations, must make the random function invariant under $C_4$. 

{\bf Orbifold}: This invariance property is achieved by shrinking the
domain $\{ x\}$ of definition of the random function $f^{\rm random}(x)$ from the square to the  $(\pi/2)$-sector, that is, to the points of the orbifold introduced above. The functional values on the square follow by rotations from $C_4$. 
Moreover the twofold Fourier basis of $f^{\rm random}(x)$ must be restricted to its subbasis invariant under $C_4$. The orbifold becomes the fundamental domain under a crystallographic space group, named  the asymmetric unit.

For cosmic  topology with curvature zero on a  finite closed Euclidean manifold  see \cite{AU08}.

\section*{Appendix B: Representation for products of Weyl reflections.}
\label{AppB}

From \cite{KR10} we recall:\\
The map of a unit vector $x$ from Caresian to  $SU(2,C)$ coordinates  $u(x)$ is given by 
\begin{equation}
\label{B1}
x=(x_0,x_1,x_2,x_3) \rightarrow 
u(x)= \left[
\begin{array}{ll}
 x_0-ix_3& -x_2-ix_1\\
x_2-ix_1& x_0+ix_3\\
\end{array}
\right].
\end{equation}
The action $T_g$ of the rotation group  $SO(4,R) \sim ((SU^l(2,C) \times SU^r(2,C))/Z_2$ with elements $g=(g_l, g_r)$
on a Wigner polynomial is, by use of representations of $SU(2,C)$ \cite{ED57}, 
\begin{equation}
 \label{B2}
(T_{(g_l,g_r)} D^j_{m_1,m_2})(u)= D^j_{m_1,m_2}(g_l^{-1}ug_r)=
\sum_{m_1'm_2'} D^j_{m_1',m_2'}(u) D^j_{m_1,m_1'}(g_l^{-1})D^j_{m_2',m_2}(g_r).
\end{equation}
For two  Weyl reflection operators with Weyl unit vectors $\{a_i, a_j\}$ define with eq. \ref{B1}
\begin{equation}
 \label{B3}
v_i:=u(a_i),\; v_j:=u(a_j).
\end{equation}
Then the rotation operator for the product $(W_iW_j)$ in terms of eq. \ref{B2}
is given by 

\begin{equation}
 \label{B4}
T_{(W_iW_l)}= T_{(v_iv_j^{-1}, v_i^{-1} v_j)}. 
\end{equation}

\end{document}